\documentstyle[pre,aps]{revtex}

\begin{document}
\title{Self-similar transport processes in a two-dimensional realization
of multiscale magnetic field turbulence}
\author{Francesco Chiaravalloti}
\address{Dipartimento di Fisica, Universit\`a della Calabria,
 I-87036 Arcavacata di Rende, Italy}
\author{Alexander V. Milovanov$^*$}
\address{Department of Physics, University of Troms\o, 9037 Troms\o, Norway}
\address{$^*$Permanent address: \\
             Department of Space Plasma Physics,
             Space Research Institute, \\
             84/32 Profsoyuznaya street,
             117997 Moscow, Russia}
\author{Gaetano Zimbardo}
\address{Dipartimento di Fisica, Universit\`a della Calabria,
 I-87036 Arcavacata di Rende, Italy}
 \address{Istituto Nazionale di Fisica della Materia, Unit\`a di Cosenza,
 I-87036 Arcavacata di Rende, Italy}
\date{\today}

\maketitle

\begin{abstract}
We present the results of a numerical investigation of charged-particle
transport across a synthesized magnetic configuration composed of a
constant homogeneous background field and a multiscale perturbation
component simulating an effect of turbulence on the microscopic
particle dynamics. Our main goal is to analyze the dispersion of
ideal test particles faced to diverse conditions in the turbulent domain.
Depending on the amplitude of the background field and the input test
particle velocity, we observe distinct transport regimes ranging from
subdiffusion of guiding centers in the limit of Hamiltonian dynamics to
random walks on a percolating fractal array and further to nearly diffusive
behavior of the mean-square particle displacement versus time. In all cases,
we find complex microscopic structure of the particle motion revealing long-time
rests and trapping phenomena, sporadically interrupted by the phases of
active cross-field propagation reminiscent of Levy-walk statistics. These
complex features persist even when the particle dispersion is diffusive. An
interpretation of the results obtained is proposed in connection with the
fractional kinetics paradigm extending the microscopic properties of transport
far beyond the conventional picture of a Brownian random motion. A calculation
of the transport exponent for random walks on a fractal lattice is advocated
from topological arguments. An intriguing indication of the topological approach
is a gap in the transport exponent separating Hamiltonian-like and fractal
random walk-like dynamics, supported through the simulation.
\end{abstract}


\maketitle

\twocolumn


\section{Introduction}

Anomalous transport phenomena in complex nonlinear dynamical
systems can often be associated with the effect of turbulence on the
microscopic particle kinetics. The multiscale interaction of nonlinear
relaxation processes and self-organization mechanisms operating in
turbulent media customarily results in a transition to a nonequilibrium
(quasi)stationary state \cite{State} dominated by long-range correlations
in space and time \cite{Treumann}. In statistically homogeneous, isotropic
systems, turbulent correlations at the (quasi)stationary state support
self-similar transport regimes that are far beyond the conventional
Gaussian diffusion. As a rule, such regimes explicitly involve the
impacts of memory, intermittency, and nonlocality; examples are
Levy-type processes \cite{Levy,Today,Basile,Denisov} and fractal
time random walks (FTRW's) \cite{Levy,FTRW,101}, dating back to
Mandelbrot and Van Ness fractional Brownian motion \cite{Ness,Feder}.

Levy processes incorporate bursty dynamics
with almost instant multiscale jumps often referred
to as ``flights" \cite{Today}. The defining feature of Levy
flights is a heavy tailed power-law jump length distribution
leading to infinite second moments of the corresponding probability
density. The problem of infinite moments may be avoided by replacing
Levy flights with Levy walks through a spatiotemporal coupling posing
inertia restrictions over the bursty events \cite{Avoid,Coupling}.
Levy walks provide suitable probabilistic model of superdiffusive
transport \cite{Extended} corresponding to faster-than-linear growth
of the mean-square particle displacement versus time. In contrast
to Levy flights, FTRW's are continuous random walk processes without
identifiable jumps. Fractal time random walk statistics relies on a
power-law distribution of waiting times between consequent steps of
the motion \cite{Sokolov}. FTRW's account for long-time particle
rests in the turbulent medium and customarily result in
slower-than-linear evolution of the tracer dispersion.
A comprehension of the essential role played by FTRW and Levy
statistics in the microscopic description of turbulence has led
to a formulation of the fractional $-$ ``strange" $-$ kinetics
\cite{Nature} reviewed in Refs. \cite{Klafter,Reports}.

Near a (quasi)stationary state, the competition between
fractal time and Levy walk regimes produces a self-similar
trajectory with the variance which grows with time raised
to a power between 0 and 2:
\begin{equation}
\left< {\bf r} ^2 (t) \right> \sim 2 {\cal {D}} t ^{\mu}, \ \ \
0\leq\mu\leq 2. \eqnum{1}
\end{equation}
The quantity ${\cal {D}}$ in Eq. (1) has the sense
of a generalized transport coefficient and is measured
in units cm$^{2}$ s$^{-\mu}$. One distinguishes between
subdiffusive ($\mu < 1$) and superdiffusive ($\mu > 1$)
processes: The former are dominated by long-time rests,
the latter, by ``active" migration regimes such as Levy
walks. The property of self-similarity appears in fractal
structure of the trajectories over a broad range of spatial
scales. The key parameter describing the topology of the
particle track in the real space is the fractal dimension
of the motion, $d_w\geq 1$ \cite{Gefen}. The value of $d_w$
is related to the transport exponent $\mu$ in Eq. (1) via
\begin{equation}
\mu = 2 / d_w, \ \ \ d_w\geq 1.
\eqnum{2}
\end{equation}
Equation (2) holds both FTRW's and flights \cite{Klafter}. The
transport regimes included in Eqs. (1) and (2) range from
ballistic motion ($d_w = 1,\mu = 2$) to the confinement ($d_w =
\infty,\mu = 0$). Diffusive processes correspond to $d_w = 2$ and
$\mu = 1$. A realization of a trajectory having $d_w = 2$ is
provided by the conventional Brownian random walk model
\cite{Feder} showing uncorrelated behavior on all time and spatial
scales \cite{Klafter}. Note that the Brownian random motion on a
plane covers densely everywhere the ambient two-dimensional space
as time $t\rightarrow\infty$ \cite{ZZ}. In case of subdiffusion
($\mu < 1$), the trajectory covers the plane with excess density
when compared to the random Brownian counterpart, meaning $d_w >
2$ in Eq. (2). The phenomenon is generally agreed to be due to
multiscale memory effects which make the tracer return more often
to the points already visited to time $t$. In contrast to
subdiffusion, superdiffusive regimes ($\mu > 1$) suppress the
returns, enabling ``fast" tracks with the ``low" dimension $d_w <
2$. An inherent drive for superdiffusion is often associated with
spatiotemporal nonlocality \cite{Klafter}, though its detailed
comprehension is far from being complete \cite{Sokolov}.

In many cases, the subdiffusive ($\mu < 1$) behavior in Eq. (1) is
supported by a self-organized concentration of turbulent transport
processes on low-dimensional (fractal) arrays such as percolating
lattices. [The term ``percolating" is synonymous with ``infinite
connected," meaning a structure which stretches to arbitrary long
(macroscopic) scales. In what follows, ``connected" is understood
in a somewhat restrictive sense ``path-connected:" An arbitrary
point found on a path-connected fractal set can propagate to
another one along a continuous trajectory which lies everywhere on
this set. The property of path-connectedness makes it possible to
consider the fractal distribution as a {\it single} topological
object.] Given a self-similar percolating fractal lattice with the
Hausdorff dimension $d_f\geq 1$, one finds the mean-square
displacement of the walker after time $t$ to be \cite{Gefen}
\begin{equation}
\left< {\bf r} ^2 (t) \right> \sim 2 {\cal {D}} t ^{2 / (2 +
\theta)}. \eqnum{3}
\end{equation}
The dispersion in Eq. (3) corresponds to the exponent $\mu = 2 /
(2 + \theta)$ and the fractal dimension of the motion $d_w = 2 +
\theta$. The quantity $\theta$ is the index of connectivity of the
fractal, which simultaneously appears in the probability to return
to the starting point $p (t)\propto t^{- d_f / (2+\theta)}$
\cite{OP85}. The index of connectivity describes intrinsic
topological features of fractal objects \cite{PRE00} and observes
remarkable invariance properties \cite{Uspehi}. For path-connected
fractal distributions, the value of $\theta\geq 0$ (as opposed to
disconnected fractals having $\theta < 0$) \cite{Uspehi}. In the
limit of Euclidean (nonfractal) geometry, the index of
connectivity $\theta \rightarrow 0$, leading to the conventional,
Einstein relation $\left< {\bf r} ^2 (t) \right> \sim 2 {\cal {D}}
t$ for the particle diffusion. Together with the Hausdorff
dimension $d_f$, the index of connectivity $\theta$ defines the
so-called spectral (or fracton) dimension $d_s = 2d_f / (2+\theta)
= 2d_f / d_w = \mu d_f$ \cite{Naka}. The value of $d_s$ determines
the effective (fractional) number of degrees of freedom on a
fractal geometry \cite{Georges}. This fractional number is
manifest in the scaling $p (t)\propto t^{- d_s / 2}$, associated
with the return probability $p (t)$ in a fractional ``Euclidean"
space enabling $d_s$ ``orthogonal" directions \cite{Uspehi}. In
view of $\theta\geq 0$, the value of the spectral dimension cannot
exceed its Hausdorff counterpart as soon as the property of
path-connectedness applies, $d_s \leq d_f$.

Physical realizations of strange transport processes (both
subdiffusive and superdiffusive) encompass fluid \cite{Fluid} and
electrostatic drift-wave turbulence \cite{EMflow,Naulin}, chaotic
flows generated by multipoint vortices \cite{Jets}, self-avoiding
random walks \cite{Grassberger}, disordered solid materials
\cite{PRB01}, astrophysical \cite{Gaetano} and laboratory
\cite{Plasma} plasma, including highly intermittent, self-similar
regimes measured in the edge and scrape-off layer region of fusion
devices \cite{Carreras}. In cosmic electrodynamics, unconventional
statistical properties of transport bridging to the strange
kinetics paradigm have been speculated for the solar photosphere
\cite{PhysFluids}, the Earth's dayside magnetopause
\cite{MZ95,Greco2003}, and the Earth's distant magnetotail
\cite{Greco2000,JGR01}.

In this paper, we analyze self-similar transport processes in a
synthesized magnetic configuration composed of a constant
homogeneous background field and a multiscale perturbation
component simulating a marginal impact of turbulence on the
behavior of test-particles. The physics included in the
consideration below ranges from laboratory (fusion) realizations
to typical conditions in space plasmas, specific to charged
particle (ion) dynamics in turbulent magnetotail-like current
sheets. Our study includes both numerical and theoretical
counterparts. The details of the numerical magnetic field model
along with the microscopic equations of motion are given in Sec.
II. The basic transport regimes {\it a priori} speculated for the
turbulent system concerned are addressed in Sec. III. Numerical
simulation results for the particles faced to diverse conditions
in the turbulent domain are presented in Sec. IV. We end with a
summary in Sec. V.

\section{Magnetic Field Model}

Setting $(x,y,z)$ to be Cartesian coordinates
in a three-dimensional Euclidean space, we define the
magnetic field configuration through $B_x = 0$, $B_y = 0$,
and $B_z = B$, where
\begin{equation}
B\, \equiv\, B_0\, +\, \delta B ({\bf r}),
\ \ \ {\bf r}\equiv (x, y, 0),
\eqnum{4}
\end{equation}
$B_0$ is a constant homogeneous background field, $\delta B ({\bf
r})$ is a magnetostatic perturbation generated across the
horizontal $(x y)$ plane, and ${\bf B} = (0, 0, B)$ is the
magnetic field vector looking everywhere in the direction normal
to the plane, $z$. The existence of a zero-frequency,
magnetostatic perturbation mode in a magnetized plasma was shown
in Ref. \cite{Ohkawa}. This mode is reminiscent of the
zero-frequency electrostatic convective-cell (vortex) mode
\cite{Taylor} in two-dimensional plasma. In particular, the two
modes share scaling properties of the cross-field transport
coefficient above the classical collisional limit \cite{Ohkawa}.
The behavior of extremely low-frequency perturbation modes in an
unmagnetized plasma have been discussed by Ginzburg and Ruhadze
\cite{Ginzburg}. The perturbation component in Eq. (4) is assumed
in the form
\begin{equation}
\delta B ({\bf r})\, =\, \sum\limits _{{\bf k}}\, \delta B _{{\bf k}}
\exp [i {\bf k}\cdot {\bf r} + i\varphi _{{\bf k}}],
\eqnum{5}
\end{equation}
where $\delta B _{{\bf k}}$ is the Fourier amplitude of the
magnetostatic mode with wave vector ${\bf k} = (k_x, k_y, 0)$, and
$\{ \varphi _{{\bf k}} \}$ are random phases posing $-$ in the
limit of statistical description $-$ the reflection symmetry
$\delta B ({\bf r}) \leftrightarrow -\delta B ({\bf r})$. This
mirrors the inherent parity properties of the multiscale magnetic
field turbulence {\it in situ} measured in the Earth's magnetotail
current sheet \cite{Hoshino}. Owing to the reflection symmetry,
the separatrix $B = B_0$ corresponding to the zero-set of the
perturbation component $\delta B ({\bf r}) = 0$ contains a
path-connected percolating subset \cite{PRE00} enabling field-line
and magnetized charged-particle transport at the macroscopic
scales \cite{Isi}. The amplitudes $\delta B _{{\bf k}}$ introduced
in the Fourier expansion in Eq. (4) are further defined through an
algebraic function
\begin{equation}
\delta B _{{\bf k}}\, =\, \frac
{2\pi A / L}{(k^2 L ^2\, +\, 1) ^{(\alpha + 1)/4}},
\eqnum{6}
\end{equation}
where $k = |{\bf k}|$ is the absolute value of the wave vector
${\bf k}$, $L$ is the side of the square, double periodic simulation
box, $A$ is the normalization constant, and $\alpha$ is the slope of
the Fourier power-law energy density spectrum, $P (k) \sim k^{-\alpha}$.
Equations (5) and (6) define the so-called {\it fractional Brownian
surfaces} \cite{Feder}, synthesized fractal objects incorporating generic
features of percolation in irregular media \cite{Isi}. [Here we ignore
nonlinear effects like mode coupling and build-up of correlations which
may destroy the wave-like features in Eq. (5). The role of nonlinearity
has recently been emphasized in Refs. \cite{Naulin} and \cite{Garcia}
in connection with the formation of coherent vortical structures in
low-$\beta$ plasmas, dominating the transport on individual flux-surfaces.
In this study, we focus on the impact of fractal behavior on
the microscopic particle dynamics, already present in the
generic realization in Eqs. (5) and (6).\,]

In the numerical model, the components $k_x$ and $k_y$ are settled on the
grid $k_x = 2\pi n_x / L$ and $k_y = 2\pi n_y / L$, with integer valued
$n_x$ and $n_y$; these satisfy $n^2 _{\min} \leq n_x^2 + n_y^2 \leq
n^2 _{\max}$. The ${\bf k}$ space is thereby a circular corona stretching
from $k_{\min} = 2\pi n _{\min} / L$ to $k_{\max} = 2\pi n _{\max} / L$.
For the runs presented below, $n _{\min} = 4$ and $n _{\max} = 80$.
Accordingly, the turbulence wavelengths $\lambda = 2\pi / k$ range
from $\lambda _{\min} = L / 80$ to $\lambda _{\max} = L / 4$. The
shortest wavelength $\lambda _{\min}$ determines the finest scale
of the inhomogeneities present, $a\sim\lambda _{\min}$. The longest
wavelength $\lambda _{\max}$, in its turn, simulates the macroscopic
turbulence correlation length, $\xi\sim\lambda _{\max}$. The ratio
$\xi / a \sim 20$. This mimics the properties of the turbulence
in the Earth's stretched and thinned magnetotail near a marginal
(quasi)stationary state \cite{JGR01}. The number of independent
Fourier modes in Eq. (5) is, by order of magnitude,
$N \sim \pi (n _{\max}^2 - n _{\min} ^2) / 2 \sim 10 ^4$. A zero
cut $\delta B ({\bf r}) = 0$ of the fractional Brownian surface
in Eq.\ (5) is illustrated in Fig. \ref{fig1}.

The equation of motion for a particle of mass $m$ and
charge $q$, migrating across the magnetic field ${\bf B}$, reads
\begin{equation}
m\frac{d {\bf v}}{dt}\, =\, q\frac{{\bf v}\times {\bf B}}{c},
\eqnum{7}
\end{equation}
where ${\bf v} = d {\bf r} / dt = (dx/dt, dy/dt, 0)$ is the
velocity at point ${\bf r} = (x, y, 0)$ at time $t$. Denote $\rho
\sim v m c / q B$ to be the particle Larmor radius in the magnetic
field ${\bf B}$. (Here $v = |{\bf v}|$ is the absolute value of
the velocity vector ${\bf v}$.) Let us further introduce
dimensionless parameters $\ell \sim \rho / a$ and $b \sim \delta B
/ B_0$. The effective cross-field transport regime may be
sensitive to $\ell$ and $b$, as we now proceed to show.

\section{Basic Transport Regimes}

\subsection{Hamiltonian limit}

Assume first $\ell\rightarrow +0$ and $b\ll 1$, meaning magnetized
particles everywhere in the $(x y)$ plane. As is well known
\cite{Krall}, the dynamics of magnetized particles is adiabatic,
enabling one to rely on a guiding center $-$ drift $-$
approximation of the long-time ($t\gg mc/qB_0$) particle motion.
The guiding center drift velocity is given by \cite{Krall}
\begin{equation}
{\bf u} = \frac{mc}{2q}\frac{v^2}{B_0}\frac{{\bf B} \times \nabla
B}{B_0^2}, \eqnum{8}
\end{equation}
leading to a system of Hamiltonian equations
\begin{equation}
\frac{d x}{d t}\, =\,
\frac{\partial\Phi (x, y)}{\partial y}, \ \ \
\frac{d y}{d t}\, =\, -
\frac{\partial\Phi (x, y)}{\partial x}.
\eqnum{9}
\end{equation}
Here, $\Phi (x, y) = - (v^2 / 2 B_0 \Omega) \delta B (x, y)$ is
time independent $-$ stationary $-$ Hamiltonian, and $\Omega \sim
q B_0 / m c$ is a characteristic particle gyrofrequency. The
Hamiltonian $\Phi (x, y)$ is considered as the zero frequency
limit ($\omega \rightarrow 0$) of a time varying, single frequency
($\propto e^{i\omega t}$) Hamiltonian $\Phi (x, y, t)$. Starting
from the periodic Hamiltonian $\Phi (x, y, t)$ and turning the
perturbation frequency to zero, $\omega \rightarrow 0$, one may
demonstrate \cite{ZZ} that the main contribution to the
cross-field transport comes from a thin layer surrounding the
separatrix $B = B_0$, where the particle excursion periods along
percolating isoenergetic contours resonate with the basic period
of the field, $2\pi / \omega \rightarrow \infty$.

In many ways, the Hamiltonian dynamics in Eqs. (8) and (9)
resembles the convection of a magnetized plasma by ${\bf E} \times
{\bf B}$ drifts caused by zero-frequency electrostatic vortex
modes across the externally applied confining magnetic field
\cite{Taylor}. The ubiquitous zero frequency $-$ static $-$ limit
$\omega \rightarrow 0$ extremizes the case of ``strong" turbulence
\cite{Strong}, conventionally described by the so-called Kubo
number ${\cal {Q}} \sim u / \omega a \gg 1$, where $u = |{\bf u}|$
is the absolute value of the vector ${\bf u}$, and $a / u$ is a
typical migration time in the field. For ${\cal {Q}} \gg 1$, the
large-scale behavior of the turbulent transport coefficient in Eq.
(1) can be summarized by the power-law dependence \cite{PRE01}
\begin{equation}
{\cal {D}}\sim \frac{1}{2} a^2
\omega ^{2/d_w} {\cal {Q}} ^{d_w / (2d_w - 1)},
\eqnum{10}
\end{equation}
where the fractal dimension $d_w\geq 1$ stands for the topology of
the transport process in the real space. The factor $\omega
^{2/d_w}$ behind $a^2$ accommodates the strange ($\propto t
^{2/d_w}$) evolution of the variance $\left< {\bf r} ^2 (t)
\right>$ versus time $t$. In case of a diffusion ($d_w = 2$), the
coefficient in Eq. (10) scales with ${\cal {Q}}$ as ${\cal {Q}}
^{2/3}$. This contrasts the widely known Bohm scaling ${\cal {D}}
\propto {\cal {Q}} ^1$, which is traditionally affiliated with
charged-particle diffusion in low-frequency wave fields, since the
pioneering works in Refs. \cite{Ohkawa} and \cite{Taylor}. The
deviation from the Bohm scaling mirrors the role of percolation
properties in the guiding center picture of the cross-field
particle transport \cite{Jacques}.

As the perturbation frequency vanishes, $\omega \rightarrow 0$,
the Kubo number in Eq. (10) diverges as an inverse, ${\cal {Q}}
\rightarrow \infty$. The transport coefficient ${\cal {D}} \propto
\omega ^{2/d_w} {\cal {Q}} ^{d_w / (2d_w - 1)}$ then goes to zero
if the fractal dimension of the motion satisfies $2/d_w > d_w /
(2d_w - 1)$, yielding $1\leq d_w < 2 + \sqrt{2}$. For $d_w
\rightarrow 2 + \sqrt{2} \approx 3.41$ [to be associated with the
stationary Hamiltonian $\Phi (x, y)$ in Eq. (9)], the value of
${\cal {D}}$ saturates at
\begin{equation}
{\cal {D}} \sim a ^{\sqrt{2}} u ^{2 - \sqrt{2}} / 2
\eqnum{11}
\end{equation}
and does not depend on $\omega$. The corresponding transport law
$\left< {\bf r} ^2 (t) \right> \propto t^{\mu}$ is a {\it subdiffusion}
with the exponent
\begin{equation}
\mu = 2 - \sqrt{2} \approx 0.58 ~~~~(t\rightarrow\infty).
\eqnum{12}
\end{equation}
The subdiffusive regime in Eq. (12) is dominated by
long-time particle rests near the points of equilibrium, defined by
$\partial\Phi / \partial x = 0$ and $\partial\Phi / \partial y = 0$.
This behavior is asymptotic: $t\rightarrow\infty$. In view of Eq. (12),
the coefficient in Eq. (11) observes anomalous scaling with
the magnetic perturbation $b \ll 1$, i.e., ${\cal {D}}
\propto b ^{2 - \sqrt{2}}$.

It is instructive to emphasize that the parameter
$\mu = 2 - \sqrt{2}$ is the exact lower bound on the transport
exponent for stochastic Hamiltonian systems with $1\frac{1}{2}$
degrees of freedom \cite{PRE01}. This lower bound is determined by the
extremely long particle rests near the points of equilibrium, where the
guiding center drift velocity vanishes. The prevalence of such long
rests in the zero frequency limit suppresses the effect of topology of
the separatrix on the cross-field transport rate. This appears in the
fact that the value $\mu = 2 - \sqrt{2}$ does not depent on the
details of the geometry of the array $B = B_0$, nor on the way the
separatrix is folded in the ambient space. Such details come into
play as soon as the magnetized (adiabatic) behavior is relaxed.

\subsection{Random walks on a fractal array}

Near the points of equilibrium, the Hamiltonian
approximation in Eq. (9) is invalidated already for very small
but finite values of the particle Larmor radius, $\ell \sim \rho / a$.
Nonadiabatic effects are more pronounced in magnetic configurations
with larger perturbation component, $b \sim \delta B / B_0$. The
limitation $b\ll 1$ assumed in Eqs. (8) and (9) may now be loosen to
$b\lesssim {\cal {O}}$, where ${\cal {O}}$ is a constant of the order
of 1. Even if slight ($\ell\ll 1$), nonadiabaticity is important as it
revives the particles attempting to rest at equilibrium. The phenomenon
helps the charge carriers to keep up their mobility near the points of
occasionally small magnetic gradients $\nabla B\rightarrow 0$. The
cross-field particle migration is naturally enhanced in this case,
implying $\mu \gtrsim 0.58$ in Eq. (1).

The inclusion of slight nonadiabaticity allows the
particles to pass more readily through the points of equilibrium and
thereby wander along the separatrix $B = B_0$ in an almost casual way.
A suitable approach to the problem could be found within a class of
random walk models \cite{Gefen} associated with the percolative
geomety of the set $B = B_0$ \cite{PRE00}.

In the case of an isotropic, random-phased
perturbation such as the fractional Brownian surface in Eqs. (5) and
(6), the separatrix $B = B_0$ forms a percolating array which observes,
in addition to the path-connectedness, {\it fractal} properties in the
range of scales between $a\sim\lambda _{\min}$ and $\xi\sim\lambda _{\max}$
(see Refs. \cite{Feder} and \cite{Stauffer}). Here, $\xi$ has the sense
of the (upper) fractal correlation length which is known to diverge at
criticality, $\xi\rightarrow\infty$ \cite{Gefen}. For ``physical" fractals,
the value of $\xi$ is of course finite ($\xi < \infty$), though far longer
than $a$. The finiteness of $\xi$ poses an upper bound where the fractal
geometry of percolation crosses over to a statistically homogeneous
distribution, $d_s\rightarrow d_f \rightarrow 2$.

In two ambient dimensions, the percolation along the separatrix is
critical, meaning a threshold at the level $B = B_0$. Crossing the
threshold one changes the domain which contains the paths to infinity
\cite{Isi}. The formation of critical percolating fractal structures has
recently been conjectured as a generic property of the (quasi)stationary
states in complex nonlinear dynamical systems far from thermal
equilibrium \cite{Uspehi}.

The criticality character supports {\it universal}
behavior of the percolation transition, manifest in a
variety of intriguing properties such as independence of the
type of the percolation problem and of the microscopic details of
the lattice \cite{Isi}. In what follows, we are interested in the
universality of the spectral dimension $d_s$ \cite{Naka}, first
conjectured by Alexander and Orbach \cite{AO} and later addressed
in an improved form by Milovanov \cite{PRE97} who introduced the
notion of the {\it percolation constant} ${\cal {C}}$, a topological
parameter incorporating the features of connectedness of fractal
distributions. The percolation constant is defined as the smaller
(between the two possible) root to the identity \cite{PRE97}
\begin{equation}
{\cal {C}}\frac{\pi ^{{\cal {C}} / 2}}{\Gamma ({\cal {C}} / 2 + 1)} = \pi,
\eqnum{13}
\end{equation}
where the symbol $\Gamma$ denotes the Euler gamma function. The value of
${\cal {C}}$, a transcendental parameter approximately equal to $1.327 \dots$,
determines the least fractional number of degrees of freedom, enabling a particle
to reach the point at infinity through a random walk process on a self-similar
fractal geometry. A topological approach to the phenomenon of percolation leading
to the identity in Eq. (13) is discussed in some detail in Ref. \cite{Uspehi}.
In terms of the percolation constant, the universality of the spectral
dimension at the threshold is quantified by \cite{PRE97}
\begin{equation}
d_s \equiv 2d_f / (2+\theta) = {\cal {C}}\approx 1.327.
\eqnum{14}
\end{equation}
Once the number of degrees of freedom is known from Eq. (14), an
evaluation of the exponent $\mu = 2 / (2+\theta) = {\cal {C}} / d_f$
in Eq. (3) is reduced to a calculation of the Hausdorff dimension $d_f$
of the lattice on which the transport process concentrates. For critical
percolation on two-dimensional fractal arrays, the value of $d_f$ lies
within \cite{PRB02}
\begin{equation}
{\cal {C}}\leq d_f\leq {\cal {S}}\equiv \ln 8 / \ln 3 = 1.89\dots < 2,
\eqnum{15}
\end{equation}
where ${\cal {S}}\equiv \ln 8 / \ln 3$ is the Hausdorff dimension of
the square Sierpinski carpet \cite{Schroeder}, a celebrated topological
object providing the universal embedding for paths on a plane \cite{Nagltop}.
Inequality (15) derives as a condition which brings together the property
of path-connectedness and the threshold character, manifest in the fact
that both ${\cal {C}}$ and ${\cal {S}}$ pose the restrictions on the
Hausdorff dimension $d_f$. In view of $\mu = {\cal {C}} / d_f$ we
have, at the critical range,
\begin{equation}
\mu _{\min} \leq\mu\leq 1,
\eqnum{16}
\end{equation}
\begin{equation}
\mu _{\min} = {\cal {C}}/{\cal {S}} \approx 0.70.
\eqnum{17}
\end{equation}
Note a gap $\Delta\mu \approx 0.12$ between
$\mu _{\min} = {\cal {C}} / {\cal {S}} \approx 0.70$ and
the Hamiltonian bound $\mu = 2 - \sqrt{2}\approx 0.58$ in Eq.
(12). Owing to the gap, a transition from Hamiltonian dynamics to
random walks on a fractal pattern may have an abrupt character, when
a slight increase in the particle Larmor radius $\ell$ results in an
almost sudden growth (from $\approx 0.58$ up to at least $0.70$) of
the transport exponent $\mu$. This makes the Hamiltonian regime in
Eq. (12) appreciably sensible to the particle Larmor radius $\ell$
assumed in the simulation.

The exact value of the Hausdorff dimension $d_f$, specific to
the topology of fractional Brownian surfaces, has been calculated
in Ref. \cite{PRE00} from the shape of fractal isoenergetic contours
in vicinity of the separatrix $B = B_0$:
\begin{equation}
d_f = 2{\cal {C}} - 1 \approx 1.65.
\eqnum{18}
\end{equation}
The Hausdorff dimension in Eq. (18) falls well inside the range defined
by inequality  (15). Relation (18) corresponds to the exponent of the
Fourier energy density spectrum obeying $1\leq\alpha\leq 7 - 4{\cal {C}}
\approx 1.69$ \cite{PRE00}. The latter guarantees enough energy at the small
scales, validating self-similar behavior of the microscopic kinetic process.
Combining Eqs. (3), (14) and (18), one arrives at
\begin{equation}
\mu = {\cal {C}} / (2{\cal {C}} - 1) \approx 0.80.
\eqnum{19}
\end{equation}
The ensuing cross-field particle transport is a {\it
subdiffusion}, $\left< {\bf r} ^2 (t) \right> \propto t^{0.80}$.
This is enhanced when compared to the subdiffusive behavior
$\left< {\bf r} ^2 (t) \right> \propto t^{0.58}$ in the limit of
Hamiltonian dynamics quantified by Eqs. (8) and (9). The fractal
dimension of the motion equals $d_w = 2 / \mu = 2 (2{\cal {C}} -
1) / {\cal {C}} \approx 2.49$ and is in fact lower than in the
adiabatic case ($d_w = 2 + \sqrt{2} \approx 3.41$). A derivation
of the turbulent transport coefficient ${\cal {D}}$ for the
particles having small but finite Larmor radius $\ell\ll 1$
involves the issue of the branching dimension \cite{ZZ} and is
intricate somewhat \cite{MZ95}. Here we note that ${\cal {D}}$ may
not reveal a single scaling behavior, contrary to the Hamiltonian
coefficient in Eqs. (10) and (11).

A subdiffusion consistent with the exponent in Eq. (19) has
earlier been proposed for ions migrating across the turbulent
magnetotail current sheet \cite{JGR01}. The estimate in Eq. (19)
applies to the ``anomalous" time scales $a^{2+\theta}\lesssim t
\lesssim \xi ^{2+\theta}$ for which the particles walk on the
fractal. For longer times $t\gtrsim \xi ^{2 + \theta}$, the
particles cover distances generally exceeding the fractal
correlation length, $\xi$. This implies a transition to
a random walk process on an Euclidean (nonfractal)
geometry. Hence
\begin{equation}
\mu\rightarrow 1
\eqnum{20}
\end{equation}
for $t\gtrsim \xi ^{2 + \theta}$. Consequently, the
asymptotic ($t\rightarrow\infty$) behavior of the cross-field
transport process in the slightly nonadiabatic case ($\ell \ll 1\,$,
$b\lesssim {\cal {O}}$) should be {\it diffusive}. The corresponding
particle diffusion coefficient is of the order of
\begin{equation}
{\cal {D}} \sim a u / 2.
\eqnum{21}
\end{equation}
The occurrence of the asymptotic ($t\rightarrow\infty$) diffusive
regime already in the slightly nonadiabatic realization is in contrast
to the magnetized particle dynamics relying on the Hamiltonian Eqs. (9).
The latter is {\it subdiffusive} (with the exponent $\mu\approx 0.58$)
on {\it all} time scales $t\gg 1 / \Omega$, up to $t\rightarrow\infty$.
Note that the diffusion cross-over time scale for random walks on a
percolating fractal array diverges as $\xi ^{2+\theta}$ for
$\xi\rightarrow\infty$.

\subsection{Strong nonadiabaticity}

In the limit of strong ($\ell \gtrsim 1$) nonadiabaticity,
the cross-field transport processes occupy wide stochastic domains,
whose measure is comparable with that of the ambient Euclidean plane
itself. The microscopic particle dynamics thereby evolves into random
walks in two dimensions ($d_s\rightarrow d_f\rightarrow 2$), meaning
{\it diffusive} behavior $\mu \rightarrow 1$ on the time scales
$t\gg a / v$. The cross-field diffusion coefficient can
be evaluated as [cf. Eq. (21)]
\begin{equation}
{\cal {D}} \sim a v \ell / 2.
\eqnum{22}
\end{equation}
Remark that the condition $\ell \gtrsim 1$ is always satisfied
near the zero-set $\delta B ({\bf r}) = 0$ for vanishing background
component, $B_0\rightarrow 0$. The perturbation parameter diverges in
this case, $b\rightarrow\infty$, signifying a decaying role of the
percolation properties.

\section{Numerical results}

In this section, we present a selection of numerical runs
which illustrate the basic transport regimes operating in
the magnetic configuration in Eqs. (4)$-$(6). Magnetic field
models with both nonvanishing and zero background component
are investigated. All lengths are normalized to the side of
the simulation box, $L$, and all times, to the inverse of
the particle gyrofrequency, $\Omega$. The running value of
$\Omega$ refers to the local magnetic field $B$ which may appreciably
deviate from $B_0$. The particle velocity $v$ is measured in
units $\Omega L$, with typical values ranging $-$ for different
realizations $-$ from $2\times 10^{-4}$ to $5\times 10^{-4}$.
For each run, 5000 particles are injected in a random manner
throughout the simulation box, $L\times L$. Initial velocity
vectors ${\bf v}$ are randomly directed. The equation of motion
(7) is solved numerically by means of a fifth-order Runge-Kutta
integrator with adaptive step; the maximum time step is $10^{-2}$.
Depending on the strength of the background field $B_0$ and the
exact value of the injection velocity $v$, the total integration
time varies from $3\times 10^6$ to $1\times 10^7$. The accuracy of
the computation is checked by various methods, including conservation
energy verification at the end of the run, showing relative errors
of less than $10^{-4}$. The evaluation of the exponent $\mu$ is
based on running fits with a time window $\Delta t \sim 10^6$,
which is made gradually move along $t$.

\subsection{Zero background component}

A nearly diffusive regime is readily recovered for the zero
background component $B_0 = 0$ (i.e., for the infinite perturbation
parameter $b = \infty$). The details of the simulation are as follows.
The particles are injected at random with the velocities $v = 5\times
10^{-4}$. The integration is performed up to $3\times 10^6$, enabling a
rich statistics of tracking data. The value of the transport exponent
computed over the whole set of the trajectories is found to be $\mu
\approx 0.96\pm 0.05$. This value incorporates the particles trapped
on closed isoenergetic contours far from the separatrix, $B = B_0$.
Simultaneously, the dispersion of those particles initialized in close
vicinity of the zero-set $\delta B ({\bf r}) = 0$ and propagating to
longest distances across the field is not distinguished from a diffusion-type
process. Sample trajectories of such particles bringing a basic contribution
in the cross-field transport are illustrated in Figs. \ref{fig2} and \ref{fig3}.
We draw attention to the very complex microscopic structure of the particle
motion, revealing a chaotic alternation of long-time rest and sporadic
transient periods, manifest in the highly irregular, intermittent way
the trajectory is marked by the tracer. In this connection, the diffusion
appears to be an intricate balance between the time intervals when the particles
stay in traps, and the ``green-light" regimes $-$ reminiscent of Levy walk
statistics $-$ when the charge carriers actively migrate across the medium.
This intriguing picture challenges the conventional Brownian random
motion paradigm customarily associated with diffusion. We interpret
this unusual behavior as a ``strange diffusion" \cite{Uspehi} deriving
from a parity between rests and walks, the two competing counterparts
of the dynamics underlying the transport at the microscopic scales.

\subsection{Finite background component}

We turn next to a situation with a finite background field $B_0$. To
support the effect of the particle Larmor radius, we set $b\sim 2$ in
the simulation below. This choice is a suitable compromise between the
opposite extremes of $b\ll 1$ and $b = \infty$.

\subsubsection{Random walks on a fractal array}

As in the regime with zero background component, we set the
injection velocity $v$ to $5\times 10^{-4}$. Accordingly, the
integration covers time scales up to $3\times 10^6$ and repeats
the details of the run specified in subsection {\bf A.} The particle
dispersion versus time is plotted in Fig. \ref{fig4}, solid line. The
fit yields $\mu\approx 0.84\pm 0.05$. This value is clearly subdiffusive
and complies with the exponent in Eq. (19). Consequently, we consider the
estimate $\mu\approx 0.84\pm 0.05$ as an evidence for a concentration of
the cross-field transport on a percolating fractal array owing to the
departure of $B_0$ from the zero limit.

To help judge the result obtained, we launch another
run with all the same integration and test particle parameters,
but with a filter in the injection scheme. As the filter is applied,
the particles found ``too far" from the percolation level $B = B_0$
are discarded. Thus, we only integrate along the trajectories that
start in a band $B_0\pm\Delta B$, with $\Delta B$ a fraction of $B_0$.
We could thereby increase the portion of particles staying close to
the percolation level and appreciably speed up the computation. The
behavior of the particle dispersion for this run is shown in Fig.
\ref{fig4}, dashed line. While larger distances are achieved, the
exponent $\mu$ remains practically unchanged, $\mu \approx 0.86
\pm 0.05$. This observation proves that the transport, as a matter
of fact, is mostly due to the particles migrating in close vicinity
of the percolation level.

A realization of
the percolation-associated particle trajectory in the presence of
the background component is developed in Fig. \ref{fig5}. The tracer
features a tendency toward adiabatic dynamics, manifest in the periods
of a drift-like migration along the separatrix. Near the saddles of the
zero-set $\delta B ({\bf r}) = 0$ (see Fig. \ref{fig1}), the drift-like
regimes are interrupted by almost unmagnetized, meandering motion in a
close-to-separatrix layer owing to the effect of finite Larmor radius.
The observation of meandering is important as it supports the chance for
casuality enabling to consider the cross-field transport in connection with
random walk processes on percolative fractal geometry \cite{Gefen}. At the
microscopic scales, the meandering motion bears signatures reminiscent of
the intermittent dynamics found for $B_0 = 0$ (see Figs. \ref{fig2} and
\ref{fig3}). The drift, in its turn, exhibits inhomogeneous behavior with
alternating phases of fast and slow progress. These competing phases of
fast and slow migration acquire the role of the walk-and-rest statistics
as the adiabatic limit is approached. We emphasize that the ``walks" are
inherently present in the motion, even though the ensuing transport appears
to be subdiffusive. In this regard, the parity between walks and rests
previously speculated for the strange diffusion is now shifted toward rests.

\subsubsection{Back to diffusion}

With increasing velocity $v$, the mean-square displacement of the
tracer tends to a linear (diffusive) form starting from longer integration
times. For shorter times, the dispersion may still be subdiffusive. For
instance, turning $v$ from $5\times 10^{-4}$ to $1\times 10^{-3}$, we
locate the transport exponent within $\mu\approx 0.95\pm 0.05$ for the
long integration time $1\times 10^7$. As the integration time is reduced
to $3\times 10^6$, the fractal random walk-like dispersion $\mu\approx
0.84\pm 0.05$ is recovered. Runs with larger velocities reveal nearly
diffusive behavior already for the short times $3\times 10^6$. These
results demonstrate the effect of the finite fractal correlation length,
$\xi$, posed by the basic periodicity of the simulation box. We emphasize
that the periodic extension of the simulation box assumed in the simulation
truncates the percolative fractal geometry of the field beyond
$\xi\sim\lambda _{\max}$.

\subsubsection{Hamiltonian limit}

To approach the adiabatic regime in Eq. (9), we turn the injection
velocity $v$ to $2\times 10^{-4}$. Simultaneously, we maintain the
background component $B_0$ at the level corresponding to $b\sim 2$
in order to achieve a clearer comparison with the previous runs. A
longer integration time $1\times 10^7$ matching the proposed decrease
in $v$ is settled. The particle dispersion as a function of time is
summarized in Fig. \ref{fig6}.

As the velocity is set to $2\times 10^{-4}$, a pronounced subdiffusive
behavior is observed through the integration period. The fit yields $\mu
\approx 0.67\pm 0.05$, which is appreciably smaller than the values allowed
for a random walk process on a fractal geometry [see Eqs. (17) and (19)].
The estimate of $\mu \approx 0.67\pm 0.05$ is reminiscent of the exponent
$\mu\approx 0.58$ in Eq. (12), corresponding to the limit of Hamiltonian
dynamics. Some deviation between the numerical result $\mu\approx 0.67\pm
0.05$ and the extreme of $\mu \approx 0.58$ could be due to the excessively
large magnetic perturbation parameter, $b\sim 2$. This deviation may further
be shown to decrease with increasing background field, $B_0$.

Turning, gradually, the injection velocity $v$
from $5\times 10^{-4}$ to $2\times 10^{-4}$, we find an unstable
transitional behavior when the exponent $\mu$ fluctuates between
Hamiltonian-like and fractal random walk-like values. This poses
an uncertainity in $\mu$ for some intermediate values of $v$. To the
major extent, the uncertainities are bypassed as $v$ gets sufficiently
close to $2\times 10^{-4}$. We associate the observed unstable regimes
with the existence of the gap deriving from Eqs. (12) and (17). On the
contrary, no specific unstable domain is found for $v$ varying between
$5\times 10^{-4}$ and $1\times 10^{-3}$, showing a continuous turnover
to the diffusive transport.

A sample trajectory corresponding to the ``cold" particles
with velocities $v = 2\times 10^{-4}$ is plotted in Fig. \ref{fig7}.
The motion is drift-like everywhere along the isoenergetic contour.
No unmagnetized, meandering effects are recognized, meaning a saturation
of the transport process at the adiabatic regime. The drift, nevertheless,
reveals inhomogeneities resembling the ``walk-and-rest" statistics, manifest
in the phases of ``fast" and ``slow" cross-field propagation. These signatures,
already noticeable for larger injection velocity $5\times 10^{-4}$, now entirely
dominate the dynamics.

\section{Summary and Conclusions}

We have analyzed the dispersion of charged particles in a
synthesized magnetic field configuration composed of a constant
homogeneous background field and a multiscale perturbation component
simulating the effect of turbulence on the test-particle dynamics.
Already for such a simple magnetic field model, we observe complex
microscopic behavior of the charge carriers, demonstrating chaotic
alternation of periods of rest and active cross-field migration
regimes reminiscent of Levy-walk statistics. The integral process
looks like a competition between rests and walks governing the
transport properties on the microscopic scales.

In absence of the background component, the cross-field transport is
almost diffusive, showing linear growth of the mean-square particle
displacement versus time. At the microscopic level, this regime profits
from an intricate compromise between the time intervals the particles
stay in traps and the periods they actively propagate through the medium.
Such complex features are in contrast with the conventional Brownian
random motion paradigm, often associated with diffusion. With increasing
background field, the equilibrium between rests and walks $-$ the two
constituents of the motion $-$ shifts toward longer rests, leading to
a sublinear (subdiffusive) behavior of the particle dispersion. We
emphasize that the walks are inherently present in the microscopic
picture of the dynamics, though the contribution they bring to the
dispersion may be suppressed by the effect of rests. Two distinct
subdiffusive regimes have further been recognized, depending on
the injection velocity $v$ (i.e., on the particle Larmor radius,
$\ell\sim v m c / a q B$).

In the numerical model, a suitable range for the velocity $v$
is allocated $-$ in dimensionless units $-$ from $2\times 10^{-4}$
to $5\times 10^{-4}$. For the larger values of $v \sim 5\times 10^{-4}$
enabling moderate nonadiabaticity of the charge carriers, the cross-field
transport is concentrated on percolating fractal arrays associated with
the zero-set of the perturbation component, $\delta B ({\bf r}) = 0$.
In two ambient dimensions, the percolation along the zero-set is critical,
meaning a threshold at the level $B = B_0$. The value of the transport
exponent as predicted by the percolation-based model equals $\mu\approx
0.80$. This result $-$ deriving from the fundamental properties of
universality of the percolation transition $-$ is reproduced in the
simulation, $\mu\approx 0.84 \pm 0.05$. A characteristic feature of
the percolation regime is the presence of chaotic meandering motion
in a thin layer enveloping the separatrix $B = B_0$,
mixed with gyromotion and trapping.

As the velocity is turned to $v = 2\times 10^{-4}$, the
transport exponent drops to $\mu\approx 0.67 \pm 0.05$. This
behavior is reminiscent of the static limit $\mu\approx 0.58$
of stochastic Hamiltonian systems with $1\frac{1}{2}$ degrees of
freedom. The tendency toward Hamiltonian dynamics reflects the
proposed decrease in $v$ in the presence of the background field.
Some deviation between the numerical estimate of $\mu\approx 0.67
\pm 0.05$ and the extreme of $\mu\approx 0.58$ may further be shown
to be due to the excessive effect of the magnetic perturbation
introduced in the simulation. This effect decays with increasing
background field.

A main result of the analytical investigation $-$ performed
in parallel with the numerical study $-$ is that the regime
of Hamiltonian dynamics is separated from random walk processes
on a percolative fractal geometry by a gap in the transport exponent,
$\Delta\mu \approx 0.12$. In the numerical simulation, the gap appears
in an uncertain behavior of the particle dispersion for some intermediate
values of $2\times 10^{-4}\lesssim v\lesssim 5\times 10^{-4}$. Within this
interval, the exponent $\mu$ fluctuates between Hamiltonian-like ($\approx
0.67 \pm 0.05$) and fractal random walk-like ($\approx 0.84 \pm 0.05$)
dispersion. The uncertanities are generally bypassed as $v$ gets
sufficiently close to $2\times 10^{-4}$.

As the Hamiltonian regime is approached, the meandering
counterpart of the particle dynamics vanishes. In this limit, the
cross-field transport is dominated by long-time particle rests near
the points of occasionally small magnetic gradients. Posing slight
nonadiabaticity readily calls for the meandering to come into play.
The transport exponent $\mu$ then promptly returns to the fractal random
walk-like dispersion. With increasing nonadiabaticity the asymptotic
diffusive regime is recovered starting from longer integration times.
This result incorporates the effect of the finite fractal correlation
length, posed by the basic periodicity of the simulation box.


\begin{acknowledgments}
It is a pleasure to thank our colleagues E. Lazzaro, M. Lontano,
H. L. Pecseli, J. J. Rasmussen, K. Rypdal, P. Veltri, L. M.
Zelenyi, and F. Zonca for illuminating and lively discussions, and
P. Pommois for his help in the numerical work. One of the authors
(A.V.M.) gratefully acknowledges the very warm hospitality at the
University of Calabria, where this study was initiated. During his
stay in Italy, A.V.M. was supported by the grants of Italian INFM
and MURST, and by the Agenzia Spaziale Italiana, contract no.
I/R122/01. In Russia, this study was sponcored by the Science
Support Foundation, by the Foundation for Basic Research (project
03-02-16967), and by the ``School-of-Science" Grant 1739.2003.2.
Partial support was received from INTAS project 03-51-3738. In
Italy, this work was granted by the Cofin 2002 (MIUR) and by the
Center for High Performance Computing (HPCC) of the University of
Calabria (Centro di Eccellenza MIUR). The final version of this
paper was prepared at the University of Troms\o, where A.V.M.
stayed on a grant from the Research Council of Norway.
\end{acknowledgments}




\newpage

\begin{figure}
\caption{A zero cut $\delta B ({\bf r}) = 0$ of the fractional
Brownian surface in Eqs. (4) and (5), corresponding to the exponent
of the Fourier energy density spectrum $\alpha = 3/2$. Owing to the
sign parity $\delta B ({\bf r}) \leftrightarrow - \delta B ({\bf r})$,
the cut $\delta B ({\bf r}) = 0$ contains a percolating isoenergetic
contour enabling charged-particle transport across the field already
in the limit of adiabatic dynamics.}
\label{fig1}
\end{figure}

\begin{figure}
\caption{A track of a test-particle injected in close vicinity
of the zero-set $\delta B ({\bf r}) = 0$ and propagating to longest
distances present. The background component is zero everywhere, $B_0 = 0$,
corresponding to the infinite value of the perturbation parameter, $b = \infty$.
The injection velocity equals $v=5\times 10^{-4}$. Dimensionless units.}
\label{fig2}
\end{figure}

\begin{figure}
\caption{A typical test-particle trajectory underlying the ``strange
diffusion." All the same injection and magnetic realization parameters
as in Fig. \ref{fig2}. Dimensionless units.}
\label{fig3}
\end{figure}

\begin{figure}
\caption{The particle dispersion as a function of time for the finite
perturbation parameter, $b\sim 2$. The injection velocity set to $v=5\times
10^{-4}$. The integration covers time scales up to $3\times 10^6$. Both runs
with (dashed line) and without (solid line) the filter are shown. In absence
of the filter, the particles are injected everywhere at random. The filter
discards the particles found ``too far" from the percolation level.}
\label{fig4}
\end{figure}

\begin{figure}
\caption{A track of a particle faced to random walk process on a
percolative fractal geometry. Note the characteristic meandering
motion mixed with inhomogeneous drift-like migration along the
separatrix.}
\label{fig5}
\end{figure}

\begin{figure}
\caption{The particle dispersion versus time for two different values
of the injection velocity, $v=5\times 10^{-4}$ (solid line) and $v=2\times
10^{-4}$ (dashed line). Finite perturbation parameter, $b\sim 2$. The
integration is performed up to $1\times 10^7$.}
\label{fig6}
\end{figure}

\begin{figure}
\caption{A track of a nearly magnetized particle migrating along the
percolating isoenergetic contour. In contrast to the trajectory in Fig.
\ref{fig5}, no meandering-like motion is present. Dimensionless units.}
\label{fig7}
\end{figure}


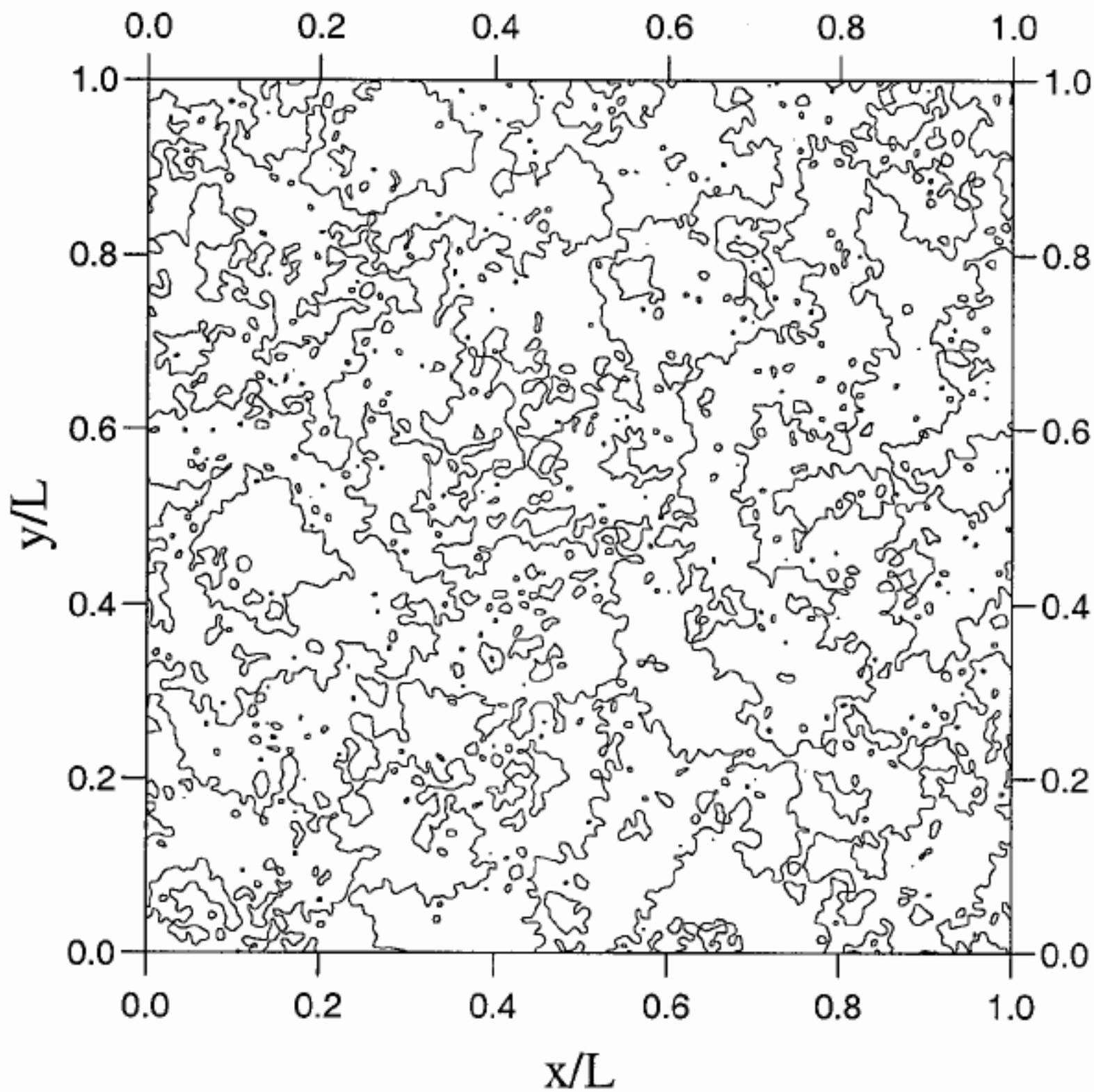
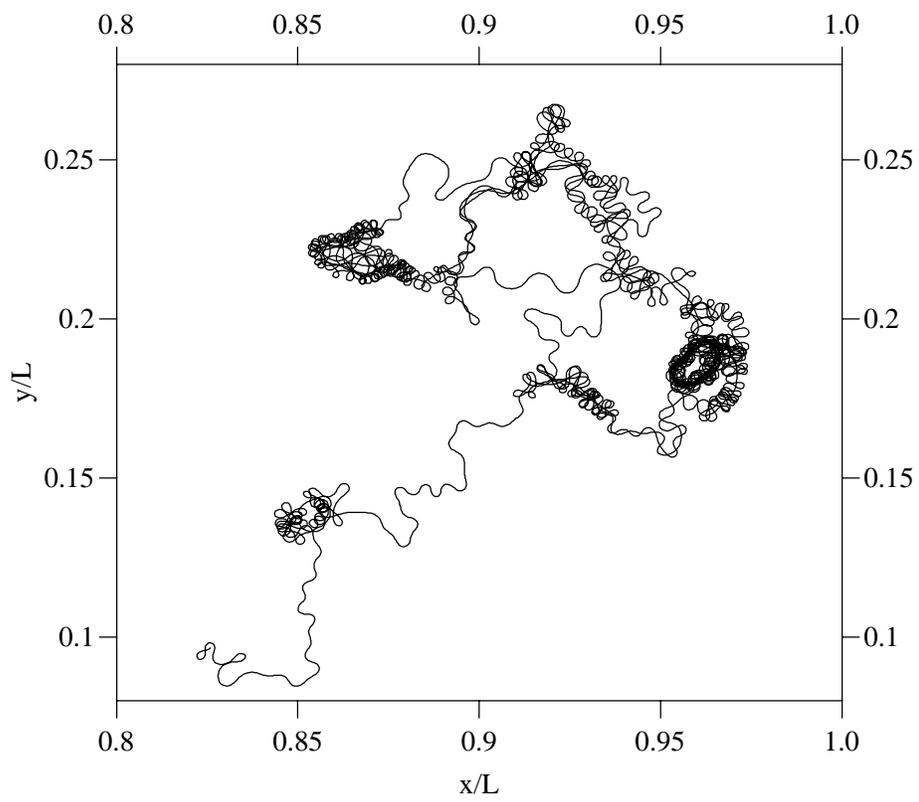
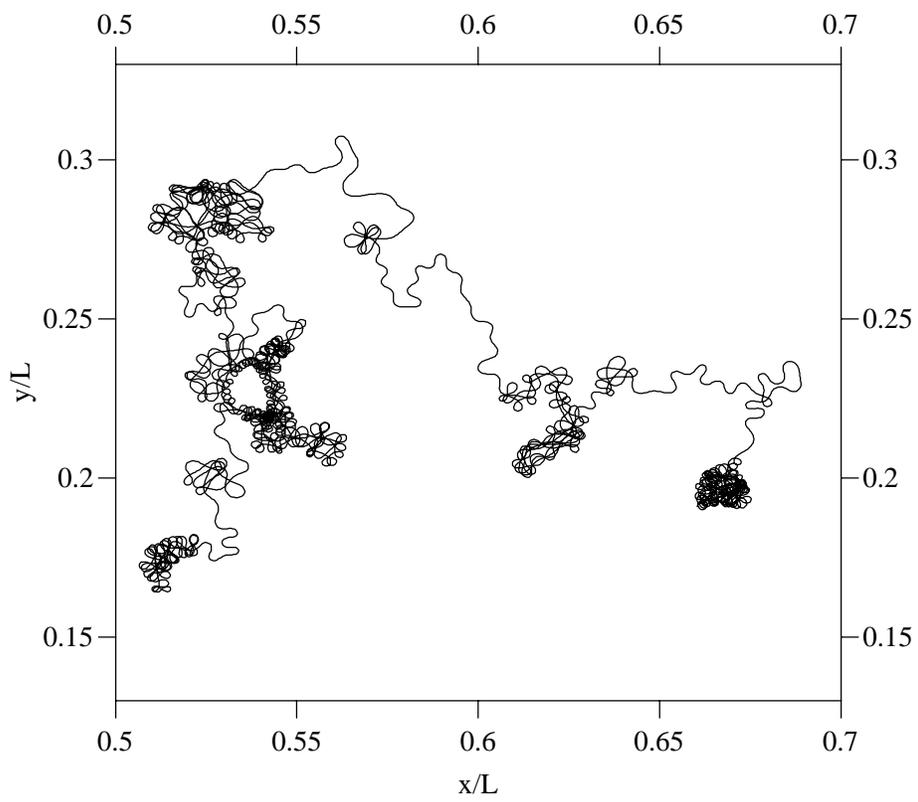
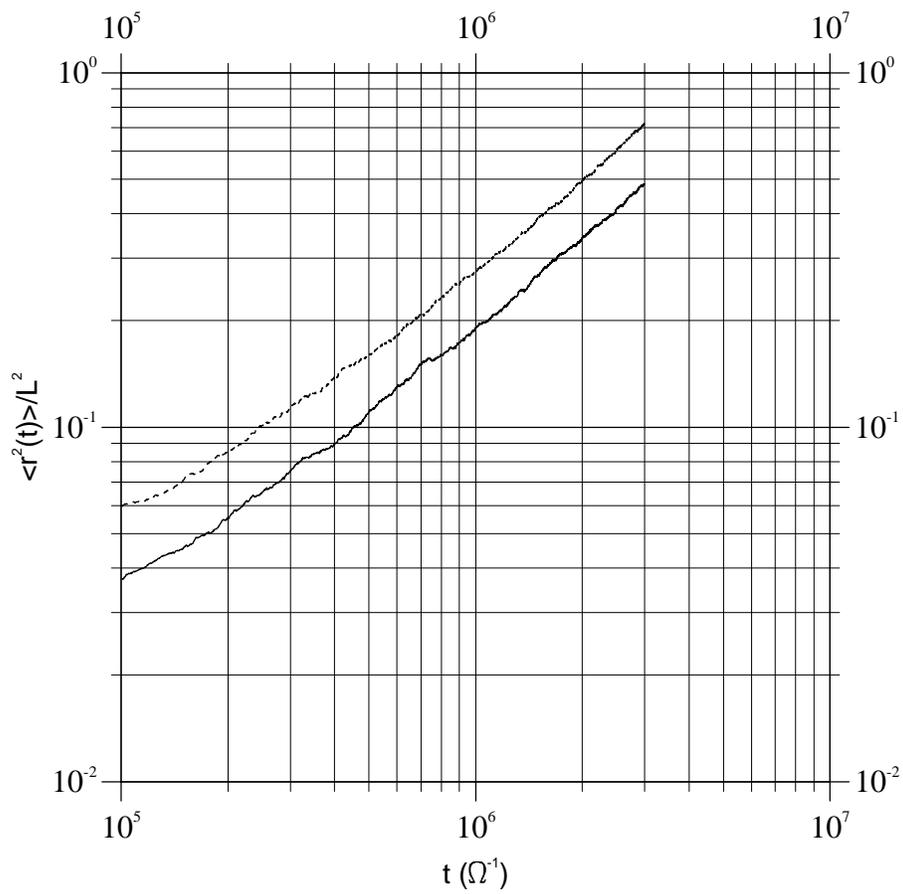
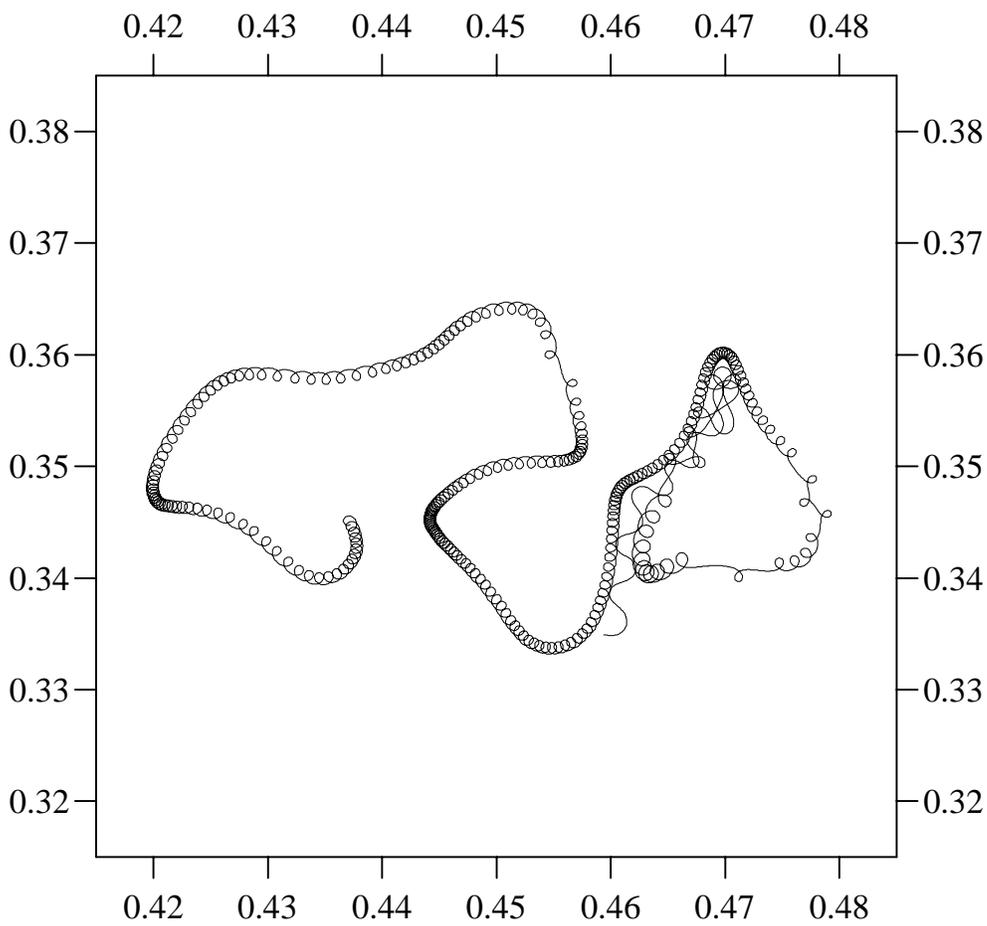
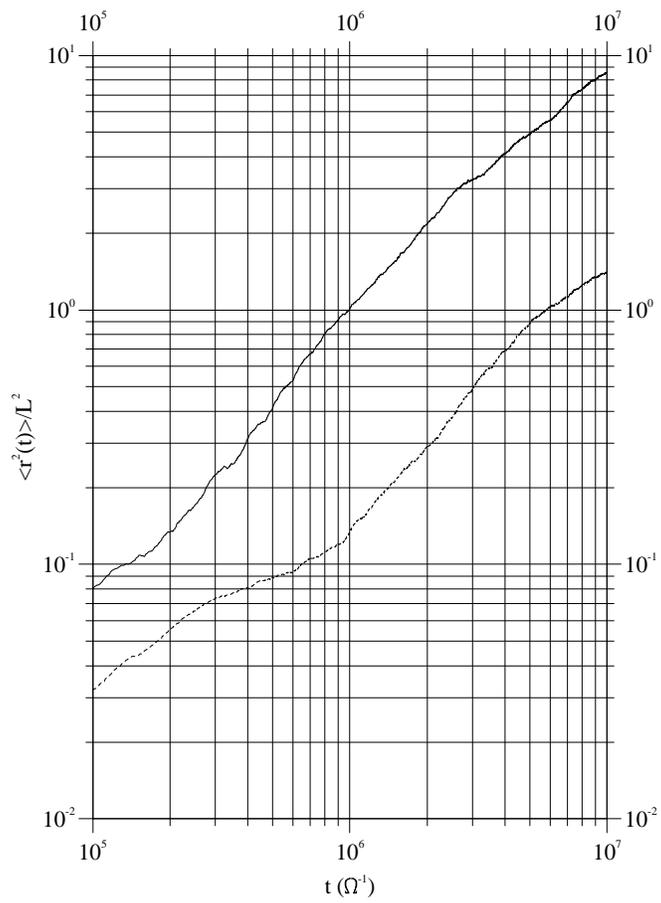
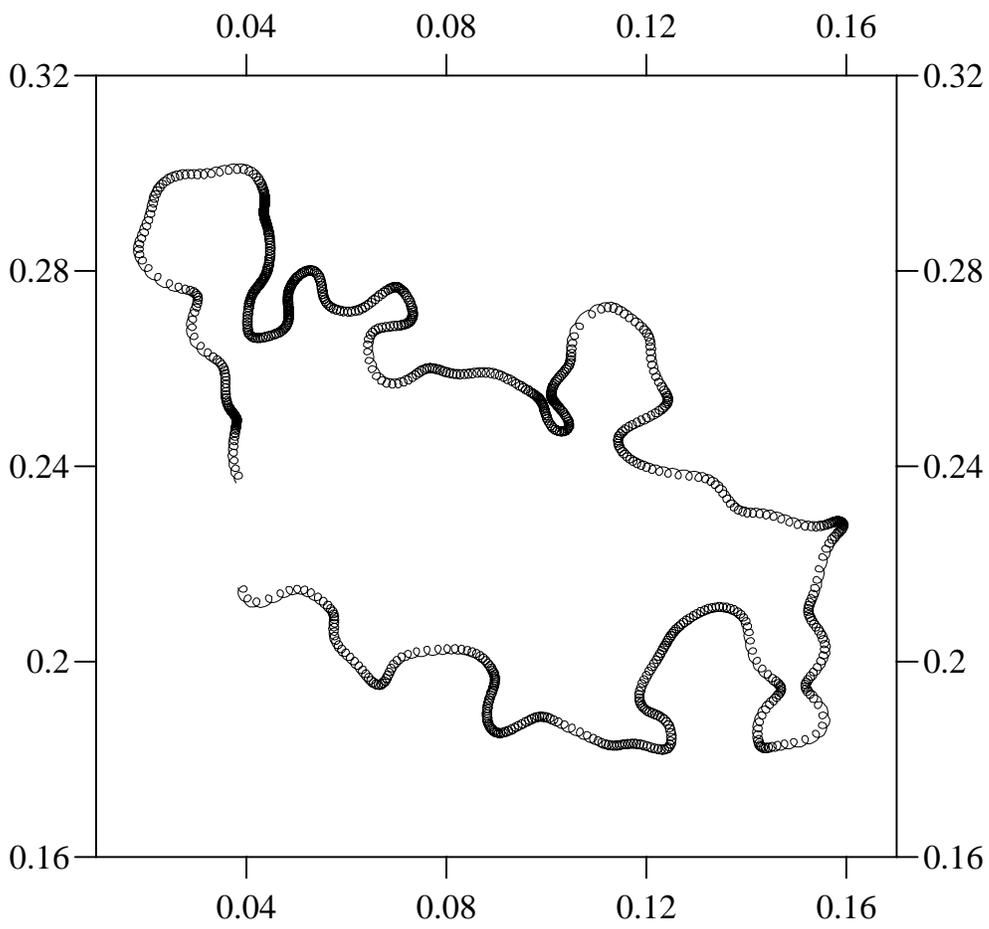
\end{document}